\documentclass[useAMS,usenatbib]{mn2e} 
\usepackage{graphicx}

\title[Environmental Dependence of Halo Growth]
{Dark Matter Halo Growth II: Diffuse Accretion and its Environmental Dependence}

\author[O Fakhouri and C-P Ma]{Onsi Fakhouri$^{1}$\thanks{E-mail: onsi@berkeley.edu, cpma@berkeley.edu} and Chung-Pei Ma$^{1}$\\
$^{1}$Department of Astronomy, 601 Campbell Hall, University of California, Berkeley, CA 94720}

\pagerange{ 
\pageref{firstpage}-- 
\pageref{lastpage}}

\usepackage{multirow}

\def \ZA {0} 
\def \ZB {0.51} 
\def \ZC {1.08} 
\def \ZD {2.07} 
\def \dz {\Delta z}
\def \dt {\Delta t}

\newcommand{\ds}{\delta_7} 
 
\newcommand{\dsfof}{\delta_{7-{\rm FOF}}}

\newcommand{\ximin}{\xi_{\rm min}}

\newcommand{\B}{\dot{M}_{\rm mer}} 
\newcommand{\C}{\dot{M}_{\rm dif}} 
\newcommand{\BZ}{d{M}_{\rm mer}/dz/M_0} 
\newcommand{\CZ}{d{M}_{\rm dif}/dz/M_0}

\newcommand{\mdott}{\dot{M}_{\rm tot}}

\usepackage{amsmath} 
\usepackage{multirow}

\begin{document}

\label{firstpage}

\maketitle 
\begin{abstract}
  Dark matter haloes in $\Lambda$CDM simulations grow by mergers with other
  haloes as well as accretion of ``diffuse'' non-halo material. We quantify
  the mass growth rates via these two processes, $\B$ and $\C$, and their
  respective dependence on the local halo environment using the $\sim
  500,000$ haloes of mass $\sim 10^{12}$ to $10^{15} M_\odot$ in the
  Millennium simulation. Adopting a local mass density parameter as a
  measure of halo environment, we find the two rates to show strong but
  {\it opposite} environmental dependence, with mergers playing an
  increasingly important role for halo growths in overdense regions while
  diffuse accretion dominating the growth in the voids.  For galaxy-scale
  haloes, these two opposite correlations largely cancel out, but a weak
  environmental dependence remains that results in a slightly lower mean
  {\it total} growth rate, and hence an earlier mean formation redshift,
  for haloes in denser environments. The mean formation redshift of
  cluster-mass haloes, on the other hand, shows no correlation with halo
  environment.  The origin of the positive correlation of $\B$ with local
  density can be traced to the sourrounding mass reservoir outside the
  virial radii of the haloes, where more than 80\% of the mass is in the
  form of resolved haloes for haloes residing in densest regions, while
  this fraction drops to $\sim 20$\% in the voids.  The negative
  correlation of $\C$ with local density, however, is not explained by the
  available diffuse mass in the reservoir outside of haloes, which is in
  fact larger in denser regions.  The non-halo component may therefore be
  partially comprised of truly diffuse dark matter particles that are
  dynamically hotter due to tidal stripping and are accreted at a
  suppressed rate in denser regions.  We also discuss the implications of
  these results for how to modify the analytic Extended Press-Schechter
  model of halo growths, which in its original form does not predict
  environmental dependence.
\end{abstract}

\section{Introduction} \label{introduction}

The bottom-up growth of structure is a hallmark of hierarchical cosmological models such as the $\Lambda$ cold dark matter ($\Lambda$CDM) model. In these universes, dark matter haloes of lower mass are expected to form earlier, on average, than massive haloes. In observations, theoretical studies, and semi-analytical modelling of galaxy formation (see \citealt{BaughReview} for a review), the mass of a halo is the key variable upon which many properties of galaxies and their host haloes depend, e.g., formation redshift, galaxy occupation number, colour, morphology, star formation rate, and stellar feedback processes.

Recent work has shown that in addition to the mass, various properties of halo formation and evolution also depend on the environment within which the haloes reside. For instance, at a fixed halo mass, older haloes have been found to be more clustered than younger haloes, and the correlation between clustering strength and formation redshift is stronger for lower mass haloes \citep{Gottlober02, ShethTormen04, Gao05, Harker06, GaoWhite07, JingSutoMo07, Wechsler06, WangMoJing07, Hahn08, MFM09}. In many of these studies, halo environment is characterised via the halo bias parameter, which is determined from the relative clustering strengths of haloes to the underlying dark matter distribution.

In Part I of this series (\citealt{FM09}; henceforth FM09), we have instead chosen to use local overdensities as a more direct measure of halo environment. The focus of the study was to quantify the environmental dependence of the merger rate of haloes. We compared a number of local overdensity variables (both including and excluding the halo mass itself), some of which had been used in earlier studies \citep{LemsonKauffmann, Harker06, WangMoJing07, Maulbetsch07, Hahn08}. The key finding in FM09 was that halo-halo mergers occur more frequently in denser regions than in voids, and that this environmental dependence is similar regardless of the merger mass ratio (e.g., minor vs major) or the descendant halo mass (galaxy- vs cluster-sized): we found mergers to occur about 2.5 times more frequently in the densest regions than in the emptiest regions. We provided an analytical formula as a function of local density to approximate this environmental trend. This expression can be used with the fit for the global mean rate of \cite{FM08} (henceforth FM08) to predict halo merger rates as a function of descendant mass, progenitor mass ratio, redshift, and environment over a wide range of parameter space.

In this paper, we build on our earlier merger rate studies by investigating
the mass growth of haloes, and its environmental dependence, via two
sources: growth from mergers with other haloes (a quantity closely related
to the results of FM09), and growth from accretion of non-halo material,
which we will refer to as "diffuse" accretion. Due to the finite resolution
of the simulations, we expect a portion of this diffuse component to be
comprised of unresolved haloes, which, in a higher-resolution simulation,
should largely follow the merger physics and scaling laws of their higher
mass counterparts in our earlier studies. The diffuse non-halo component,
however, can in principal also contain truly diffuse dark matter particles
that either were tidally stripped from existing haloes or were never
gravitationally bound to any haloes.

As reported below, we find that diffuse accretion plays an important role
in contributing to halo mass growth in the Millennium simulation. Moreover,
we will show that this growth component correlates with the local halo
environment in an opposite way from the component due to mergers, with
diffuse accretion playing an increasingly important role in halo growth in
the voids, and mergers playing a more important role in the densest
regions. This difference suggests that the diffuse component is not simply
an extension of the resolved haloes down to lower masses, but rather that
there is an intrinsic difference between the two components that results in
the opposite environmental trends. An implication of this result is that
Milky-Way size galaxies in voids and those that reside near massive
clusters may have statistically distinguishable formation history, where
the galaxies in voids acquire their baryons more quiescently via diffuse
accretion, while those in dense regions assemble their baryons mainly via
mergers.  Such environmental effect can show up in galaxy properties
such as the star formation rates, colors, and morphologies.

The environmental dependence of halo growths reported in this paper also
has far-reaching implications for the much-used analytic theories for halo
growth such as the Extended Press-Schechter (EPS) and excursion set models
\citep{PS74, BondEPS,LC93}. These models assume that all dark matter
particles reside in haloes, and halo growths depend only on mass and not
environment. As we will elaborate on below, both assumptions are too
simplistic and must be modified to account for the results from numerical
simulations.

This paper is organised as follows. \S\ref{MillenniumSim} summarises the
various definitions of halo mass and environment used in our analysis, as
well as the means by which we extract halo merger trees from the public
data in the Millennium simulation. In Sec.~3 we discuss how the two mass
growth rates due to mergers and diffuse accretion, $\B$ and $\C$, are
defined and computed. The distributions of the rates for the $\sim 500,000$
haloes at $z=0$ and their redshift evolution are presented. The
environmental dependence of halo growths is analysed in Sec.~4. The
correlations of four halo properties with the local density parameter are
investigated: the mass growth rates $\B$ and $\C$ (\S~4.1), the fraction of
a halo's final mass gained via mergers vs. diffuse accretion (\S~4.2), the
formation redshift $z_f$ (\S~4.3), and the composition of the surrounding
mass reservoir outside of the virial radii of the haloes
(\S~4.4). \S\ref{Sojourners} discusses a test that we have performed to
verify that the majority of haloes reside in a similar environmental region
(e.g. overdense or underdense) throughout their lifetimes. In Sec.~5 we
investigate further the nature of the ``diffuse'' component by varying the
mass threshold used to define $\B$ vs $\C$. We then discuss the
implications of our results for the analytic EPS model of halo growth,
which is entirely independent of environment in its basic form and
therefore must be modified to account for the various environmental trends
reported in \S~4.

\section{Merger Trees and Halo Environment in the Millennium Simulation} \label{Definitions} 
\begin{table*}
	\centering 
	\begin{tabular}
		{l|ccccc} & \multicolumn{5}{c}{Mass Percentile (Masses in units of $10^{12} M_\odot$)} \\
		Redshift & 0-40\% & 40-70\% & 70-90\% & 90-99\% & 99-100\% \\
		\hline \hline $z=\ZA\,(\dz=0.06,\dt=0.83\,{\rm Gyr})$ & $1.2-2.1$ & $2.1-4.5$ & $4.5-14$ & $14-110$ & $>110$ \\
		\hline $z=\ZB\,(\dz=0.06,\dt=0.38\,{\rm Gyr})$ & $1.2-2.0$ & $2.0-4.1$ & $4.1-12$ & $12-74$ & $>74$ \\
		\hline $z=\ZC\,(\dz=0.09,\dt=0.34\,{\rm Gyr})$ & $1.2-1.9$ & $1.9-3.7$ & $3.7-9.5$ & $9.5-48$ & $>48$ \\
		\hline $z=\ZD\,(\dz=0.17,\dt=0.25\,{\rm Gyr})$ & $1.2-1.8$ & $1.8-3.0$ & $3.0-6.5$ & $6.5-24$ & $>24$ \\
		\hline 
	\end{tabular}
	\caption{Mass bins used in this paper for redshifts $\ZA,\ZB, \ZC$, and $\ZD$. The bins are computed assuming fixed mass-percentile bins (header row) and the mass boundaries are computed using the prescribed percentile. The high mass bins extend out to $5.2\times10^{15},3\times10^{15},1.3\times 10^{15}$, and $4.4\times10^{14} M_\odot$ for $z=\ZA,\ZB,\ZC,\ZD$ respectively, though these exceptionally high mass objects are outliers and not particularly representative of the high mass range.} \label{table:MassBins} 
\end{table*}

\subsection{Merger Trees of the Millennium Haloes} \label{MillenniumSim}

The Millennium simulation \citep{Springel05} assumes a $\Lambda$CDM model with $\Omega_m=0.25$, $\Omega_b=0.045$, $\Omega_\Lambda=0.73$, $h=0.73$ and a spectral index of $n=1$ for the primordial density perturbations with normalisation $\sigma_8=0.9$ \citep{Springel05}. The dark matter N-body simulation followed the trajectories of $2160^3$ particles of mass $1.2\times10^9 M_\odot$ in a (685 Mpc)$^3$ box from redshift $z=127$ to $z=0$. A friends-of-friends group finder with a linking length of $b=0.2$ is used to identify $\sim 2\times 10^7$ dark matter haloes in the simulation down to a mass resolution of 40 particles ($\sim 4.7\times 10^{10} M_\odot$). Each FOF halo thus identified is further broken into constituent subhaloes, each with at least 20 particles, by the SUBFIND algorithm that identifies gravitationally bound substructures within the host FOF halo \citep{Springel01SUBFIND}.

Even though the Millennium public database provides a catalogue of FOF haloes at each output, it does not give the merger trees for these haloes. Instead, it provides merger trees for the {\it subhaloes}, which are constructed by connecting the subhaloes across the 64 available redshift outputs. During this construction, a decision must be made about the ancestral relations of the subhaloes since the particles in a given subhalo may go into more than a {\it single} subhalo in the subsequent output. In this case, a subhalo is chosen to be the descendent of a progenitor subhalo at an earlier output if it hosts the largest number of bound particles in the progenitor subhalo. The resulting merger tree of subhaloes can then be processed further to construct the merger tree of the FOF haloes. This construction is non-trivial due to the fragmentation of FOF haloes; this is discussed at length in FM08 and FM09.

In FM08 and FM09, we proposed a variety of post-processing algorithms to handle FOF fragmentation. Three methods were compared in FM08: {\it snip, stitch-3}, and {\it stitch-$\infty$}. The snipping method severs the ancestral relationship between halo fragments and their progenitors. This is the method commonly used in the literature and suffers from inflated merger rates due to the aberrant remerger of snipped fragments. The two stitching algorithms prevent halo fragmentation by "stitching" halo fragments together such that each FOF halo in the simulation has exactly one descendant. Stitch-$\infty$ performs this procedure whenever fragmentation occurs, whereas stitch-$3$ only reincorporates halo fragments that are destined to remerge within 3 outputs of the fragmentation event. Both stitching algorithms lower the minor merger rate as they prevent spurious remergers of halo fragments. Naturally, this reduction is strongest for stitch-$\infty$.

In FM09 we showed that the rate of halo fragmentation was a strong function of halo environment, with haloes residing in overdense regions undergoing fragmentations over three times as often as haloes in underdense regions. We showed that the choice of post-processing algorithm was important in determining the environmental dependence of halo merger rates in dense environments and introduced two new algorithms: {\it split-3} and {\it split-$\infty$}. These algorithms remove halo fragmentations by splitting progenitor haloes that fragment into multiple haloes to ensure that each halo has exactly one descendant. As with stitch-3, split-3 only splits progenitors that are split within the past 3 outputs (looking backwards towards high $z$) of the fragmentation event. Split-$\infty$ \emph{always} splits fragmenting haloes. See also \cite{Genel} who introduced a similar algorithm.

We showed in FM09 that split-$\infty$ suffered from an "unzipping" effect
as halo fragmentations propagated up the tree, breaking the self-similarity
in the merger rate, $B/n$, observed in FM08.  Split-$3$, however, yielded
very similar merger rates (within $\sim 10$\%) as stitch-$3$ (see Fig.~10
of FM09). 

We have tested the various quantities presented in this paper using all five post-processing algorithms and have indeed found the differences not to be significant enough to warrant presentating plots for all methods. We will, however, briefly discuss any relevant differences in the text. The results presented throughout this paper will use the stitch-3 algorithm used in FM08 and FM09.

\subsection{Halo Environment}

To measure each halo's local environment, we will primarily use the overdensity variable $\delta_{7-\rm{FOF}}$ presented in FM09. This quantity computes the overdensity within a sphere of radius 7 Mpc, centred on each halo in the simulation, by adding up the mass contributions made by the particles within the sphere and subtracting the FOF mass of the central halo. The database provides the dark matter density on a $256^3$ grid but not the full particle distribution itself. We have mapped this grid, which is given along a Peano-Hilbert space filling curve, to spatial coordinates and interpolated it to obtain $\delta$ at the position of each halo's centre.

We will also compare some of the results in this paper to the alternative environmental variable $\delta_7$ discussed in FM09, where $\delta_7$ is the overdensity surrounding a halo within a sphere of radius 7 Mpc (including the halo mass itself). Fig.~1 of FM09 showed that $\delta_7$ was strongly correlated with mass for haloes above $10^{14}M_\odot$ when the central object's contribution to $\delta$ began to dominate at these mass scales. This correlation was absent when the variable $\delta_{7-\rm{FOF}}$ was used. We will therefore use $\delta_{7-\rm{FOF}}$ as the environmental measure for most of the analyses below.

\subsection{Halo Mass Definitions} \label{HaloMass}

The Millennium database provides a number of mass measurements for each identified FOF halo. Two simple mass measures are $M_{FOF}$, the total mass of the particles connected to an FOF by the group finder, and $M_{SH}$, the sum of the masses of the subhaloes that constitute the FOF halo. The latter mass includes only particles that are bound by the SUBFIND algorithm; thus $M_{SH} \le M_{FOF}$ by definition.

In FM08 and FM09 we used $M_{FOF}$, the total mass of the particles that belong to an FOF halo by the group finder. \cite{Genel} have shown, however, that for the subset of haloes that are about to undergo minor mergers, the FOF mass of the lower-mass halo can be significantly higher (up to a factor of 1.4) than its subhalo mass ($M_{SH}$) due to deficiencies in the FOF halo-finding algorithm. \cite{Genel} showed that this effect is more severe for more minor mergers. In light of this study, we will present the majority of our results for both $M_{FOF}$ and $M_{SH}$ and comment on the difference. 

Since the most massive haloes at $z=0$ are more massive than the most massive haloes at higher redshifts, we will present some results below using mass bins defined by a fixed percentage of haloes rather than by a fixed absolute mass at different redshifts. Table~\ref{table:MassBins} lists the corresponding halo masses at $z=0$, 0.51, 1.08, and 2.07 for each of the five percentile bins. Note that even in the highest 1\% mass bin, there are $\sim 5000$ cluster-mass haloes available for our study.

\section{Diffuse Accretion}

\subsection{Quantifying Halo Growth via Mergers vs Diffuse Accretion} \label{QuantifyingHaloGrowth}

The merger trees of FOF haloes constructed in \citet{FM08} provide, for a descendant halo of mass $M_0$ identified at redshift $z_0$, a list of the masses of its $N_p$ progenitors at any chosen $z_1$. We label the progenitors $M_i$, with $i=1,...,N_p$, and $M_1\geq M_2\geq M_3...$ etc. The number of progenitors, $N_p$, for a given descendant can range from 1 (i.e. no mergers in that timestep), 2 (binary mergers), to high values for massive haloes. The variable $\xi$ is used to denote the mass ratio of a merging progenitor and the largest progenitor $M_1$: $\xi=M_{i>1}/M_1$.

Typically we find that the mass $M_0$ of a descendant halo is not equal to the mass in the progenitors $\sum_{i\geq1} M_i$, leading us to define 
\begin{equation}
	\Delta M=M_0-\sum_{i=1}^{N_P} M_i \,, \label{eqn:DM} 
\end{equation}
where $\Delta M$ quantifies the portion of the mass change that cannot be attributed to mergers of resolved haloes $M_i$, and $\Delta M$ can be negative. A non-zero $\Delta M$ can be due to the merging of unresolved progenitors below our minimum halo mass, accretion of dark matter that is not locked up in haloes, or mass loss processes such as tidal stripping. We collectively refer to $\Delta M$ as {\it diffuse accretion} throughout this paper.

To quantify the relative contributions to halo growth due to mergers vs diffuse accretion we define two mass growth rates for each halo of mass $M_0$ at some redshift $z_0$:

\begin{eqnarray}	
	\B &\equiv & \frac{\sum_{i\geq2} M_i}{\Delta t} \,, \nonumber\\
	\C &\equiv &\frac{\Delta M}{\Delta t} \,, \label{eqn:BC} \\
	\mdott &=& \B + \C = \frac{M_0-M_1}{\Delta t} . \nonumber	
\end{eqnarray}
where $\B$ is due to mergers and $\C$ is due to the accretion of diffuse material. Since $M_0-M_1$ is the mass difference along the thickest branch of the merger tree, the sum, $\mdott=\B + \C$, is simply the net mass growth rate of the halo. A detailed study of $\mdott$ of the haloes in the Millennium simulation is given in \citet{MFM09}; here our focus is on the separate contributions and their environmental dependence. The timestep $\Delta t$ in equation~(\ref{eqn:BC}) is the time interval corresponding to $\dz=z_1-z_0$, the redshift spacing between two Millennium outputs used to calculate the rate. We use $\dz=0.06$ for $z_0=\ZA$ ($\Delta t = 0.83$ Myrs) and $\ZB$, $\dz=0.09$ for $z_0=\ZC$, and $\dz=0.17$ for $z_0=\ZD$, following the extensive convergence tests reported in \citet{FM08}.

We also find it convenient to present the results for the mass growth rates in dimensionless units. When these occasions arise, we will use the fractional mass gain per unit redshift, $\BZ$ and $\CZ$, which are simply equal to the rates in equation~(\ref{eqn:BC}) multiplied by $\Delta t/\Delta z/M_0$.
\begin{figure*}
	\centering 
	\includegraphics{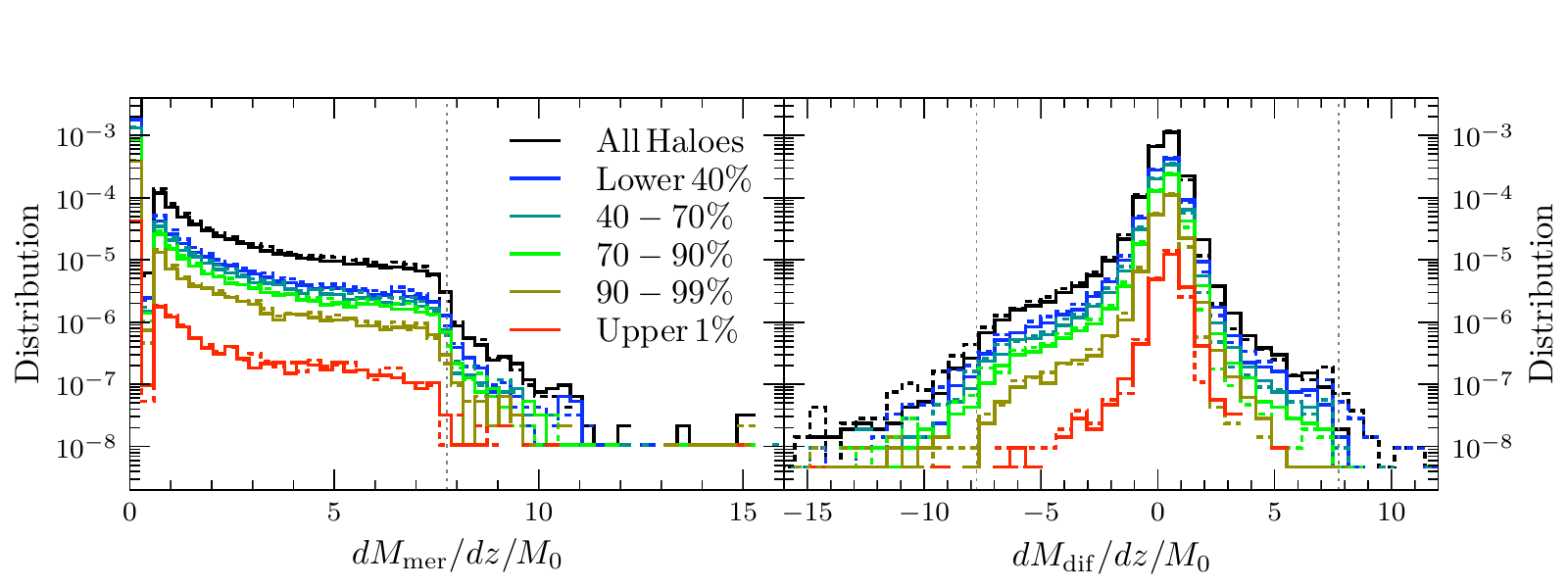} \caption{Distributions of the mass growth rates (per unit redshift) via mergers (left panel) and via diffuse accretion (right panel) for the $\sim 500,000$ haloes at $z=0$ in the Millennium simulation. In each panel, the colours denote haloes of different mass percentiles (see Table~\ref{table:MassBins}), and two definitions of halo mass are shown (dashed for $M_{FOF}$; solid for $M_{SH}$).} \label{fig:BCDistribution} 
\end{figure*}

It is important to keep in mind that the distinction between mass growth from resolvable haloes and diffuse accretion depends on the mass threshold used to define these two components. In this study, we have made the conservative choice of including only descendant haloes with more than 1000 particles (with a corresponding mass of $1.2\times 10^{12} M_\odot$) and progenitor haloes with more than 40 particles ($4.8\times 10^{10} M_\odot$). A mass ratio of $\ximin=0.04$ is therefore the resolution limit on progenitors for the most poorly resolved descendant haloes (1000 particles) in our sample. For these haloes, the mass contributed by progenitors above 40 particles is tagged as due to mergers, while any remaining mass contribution is tagged as due to diffuse accretion.

For more massive descendants, there are two natural choices for the mass threshold: one can include either all progenitors down to the 40 particle resolution limit as mergers, or all progenitors down to a fixed mass {\it ratio} of $\ximin=0.04$ as mergers (any progenitors with $\xi<\ximin$ are then counted as "diffuse" material). The former has the advantage that 40 particles correspond to a smaller $\ximin$ for more massive descendants; we will therefore have better statistics and dynamic range for more massive haloes. The latter choice has the advantage that mergers and diffuse accretion can be compared objectively across different descendant mass bins since the two components are defined with respect to the same $\ximin$ for all haloes and are, therefore, effectively equally well resolved. In Sec.~\ref{InterpretingC} we will show that the basic behaviour of the environmental dependence presented in this paper does not depend on which definition is used. We generally favour the latter definition, and will use $\ximin=0.04$ to separate mergers vs. diffuse accretion unless otherwise stated in this paper.

\subsection{Distributions of $\B$ and $\C$} \label{BCDistribution}

Fig.~\ref{fig:BCDistribution} presents the distributions of the mass growth
rates due to mergers (left panel) and diffuse accretion (right panel) at
$z=0$ for five halo mass bins. The two halo mass definitions discussed in
Sec.~2.3, $M_{FOF}$ (dashed) and $M_{SH}$ (solid), are compared.  In
constructing these plots, we have used a $\xi>\ximin$ cut: mergers with mass
ratios less than $\ximin=0.04$ are counted as diffuse accretion. 

We have also imposed a cut on the very negative and positive tails of the
distributions.  Although the majority of the haloes in the Millennium
simulation have reasonable $\BZ$ and $\CZ$, about $0.3\%$ of them more than
doubled their mass via diffuse accretion between $z=0$ and $z=0.06$, and
about $0.4\%$ lost over half of their current mass via diffuse accretion.
Our detailed examination of the merger histories of haloes in these tails
of the distributions shows that these very large values of $|\CZ|$ are
unphysical and are due primarily to complex and spurious halo
fragmentation.  The algorithm used to construct the FOF merger trees,
stitch-3, only stitches halo fragmentations that are destined to remerge
within 3 snapshots. Some fragmentations do not satisfy this condition.

This small set of haloes -- a set too small to affect the merger statistics
studied in our earlier work -- can significantly affect the mean values of
$\BZ$ and $\CZ$.  This is especially true when these mean values are
computed as a function of halo environment, since fragmentations occur more
frequently in the messier dense environments. Moreover, haloes with
artificially high growth rates have correspondingly artificially low
formation redshifts ($z_f$). These events pollute the distribution of $z_f$
and tend to underpredict the formation redshift, especially in overdense
regions. We therefore implement a cut on $\BZ$ and $\CZ$ to remove this
aberrant population. This is done by demanding that a halo never gains more
than a fraction $f_{+}=1/2$, nor lose more than a fraction $f_{-}=1/2$, of
its \emph{final} mass $M_0$ in either $\BZ$ or $\CZ$ between any two
adjacent simulation outputs along the halo's main branch (dotted vertical
lines in Fig.~\ref{fig:BCDistribution}). This is a fairly stringent cut as
it is applied to a halo's entire mass history. The resulting cut removes
$\sim 3\%$ of the haloes in our sample. Varying the values of $f_{+}$ and
$f_{-}$ in the range of $f\in[0.2,1.]$ and allowing $f_{+}\neq f_{-}$ made
insignificant changes and did not affect the overall trends reported in
this work.  A handful haloes with rates beyond these cuts remain in Fig.~
\ref{fig:BCDistribution} because the cut is applied to the halo growth across
\emph{adjacent} outputs whereas the rates shown in Fig.~
\ref{fig:BCDistribution} are computed across the three outputs between
$z=0$ and $z=0.06$ (which straddle the $z=0.02,0.04$ outputs).

In the left panel of Fig.~\ref{fig:BCDistribution}, the strong peak at
$\BZ=0$ is due to the fact that between $z=0$ and 0.06 roughly $65\%$ of
haloes have only one progenitor and, therefore, experience no mergers. The
remaining $35\%$ of haloes have a distribution that falls off smoothly out
to $\BZ\sim8.3$ (dotted vertical line). Values exceeding $8.3$ correspond
to haloes that have gained more than half their mass between $z=0.06$ and
$z=0$ (i.e., $\sum M_i/M_0/\dz>1/2/0.06=8.3$); only $\sim 0.2\%$ of haloes
are in this category.  In the right panel of Fig.~\ref{fig:BCDistribution},
the peak occurs at small positive values of $\CZ$, though a significant set
of haloes experience diffuse stripping.

\begin{figure*}
	\centering 
	\includegraphics{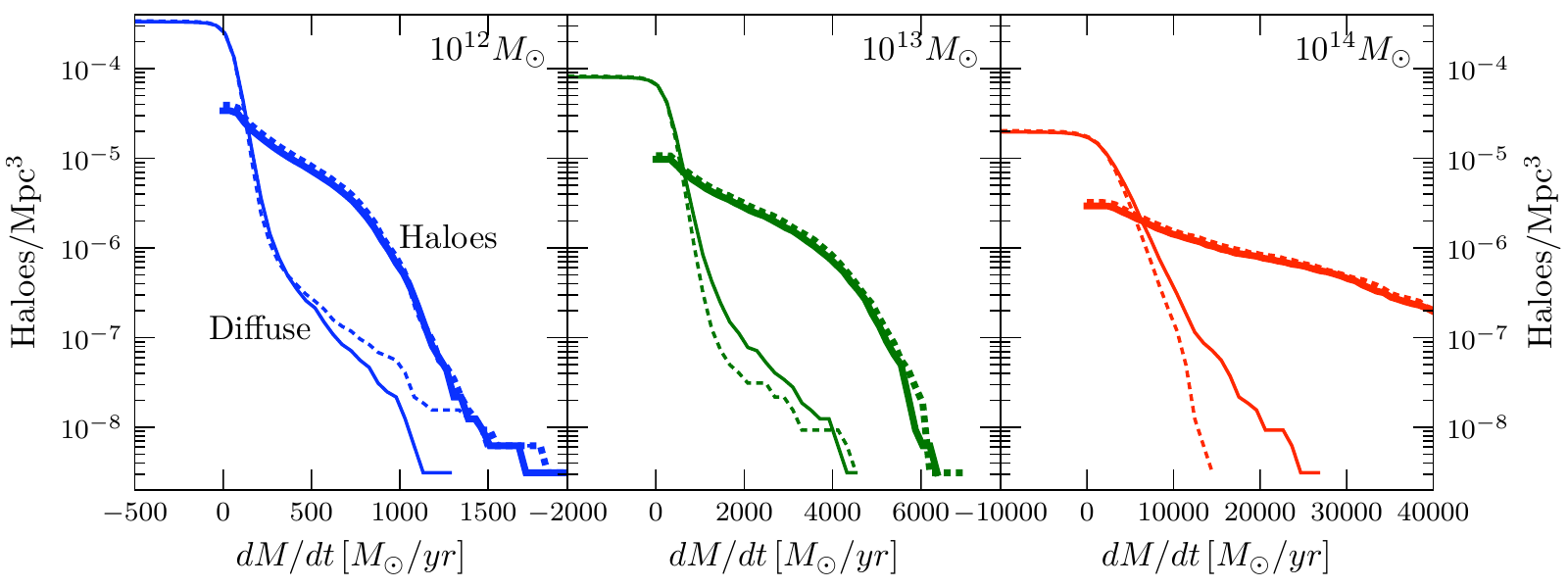} \caption{ Cumulative distributions of
          the $z=0$ mass growth rates via mergers with haloes (thick
          curves) and diffuse accretion (thin curves) for haloes in three
          mass bins (left to right panels). The axes are in physical units:
          halo number density (y-axis) and accretion rate in $M_\odot$ per
          year (x-axis). The total rate is plotted in black. The solid and
          dotted curves are for the two halo mass definitions $M_{SH}$ and
          $M_{FOF}$. The high accretion rate events are dominated by
          mergers with other haloes.} 
\label{fig:BC_Cumulative}
\end{figure*}

To complement the differential distributions presented in
Fig.~\ref{fig:BCDistribution}, we show in Fig.~\ref{fig:BC_Cumulative} the
{\it cumulative} distributions for the number density of haloes above a
given mass growth rates in units of $M_\odot$ per year. The three panels
are for three halo mass bins. The contributions from mergers (thick
curves), non-halo material (thin curve), and the sum (black curves) are
plotted separately. The curves for the total rate are identical to the
solid curves ($z=0$) in the bottom panels of Fig.~5 of \citet{MFM09}. This
figure shows that the haloes with high accretion rates in the simulation
are undergoing mergers with other haloes, while the haloes with low
accretion rates are mostly accreting non-halo mass.

\subsection{Redshift Evolution of the Mean Growth Rates}
\begin{figure*}
	\centering 
	\includegraphics{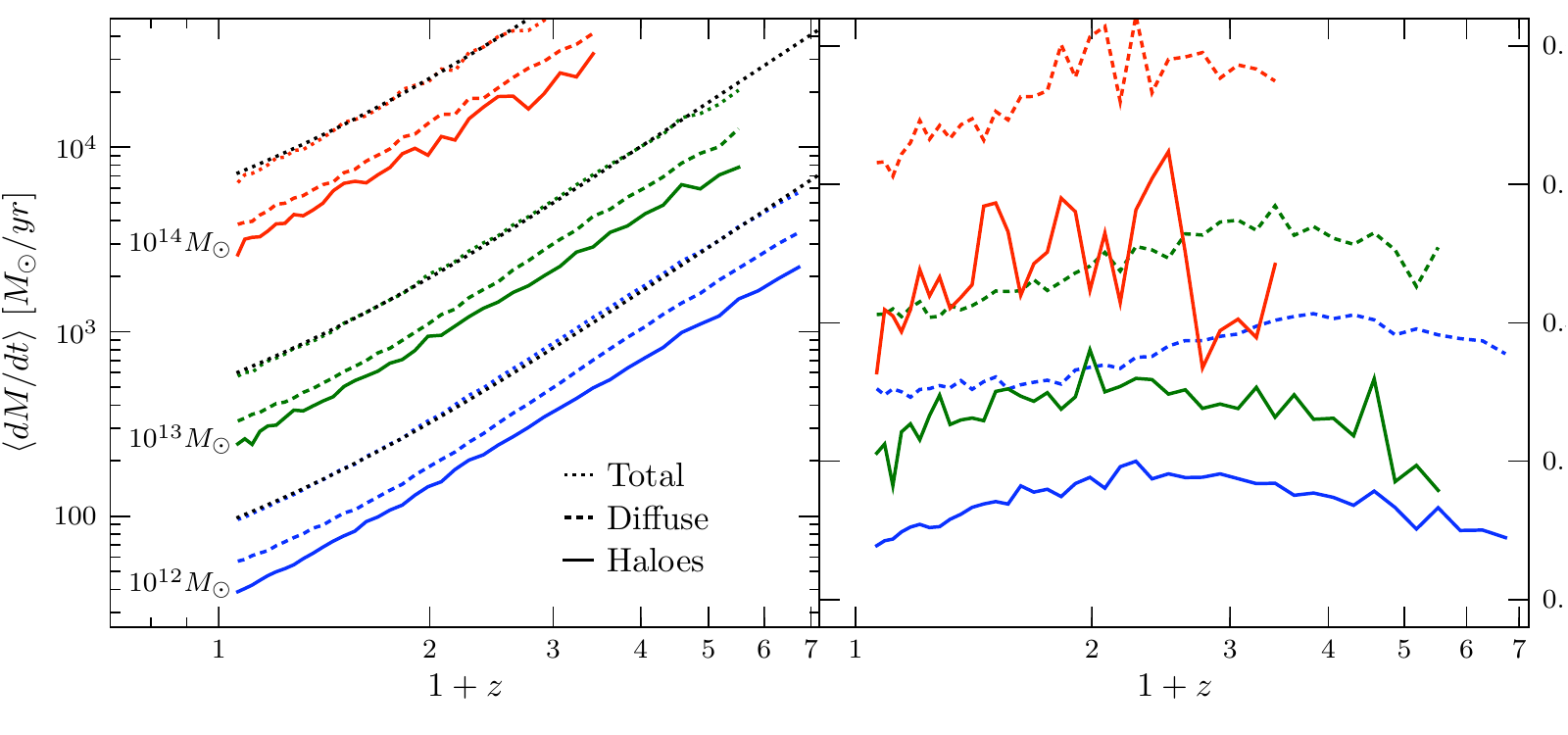} \caption{Redshift evolution of the mean halo mass growth rates due to mergers (solid) and diffuse accretion (dashed) for three mass bins: $10^{12}M_\odot$ (blue), $10^{13} M_\odot$ (green), and $10^{14} M_\odot$ (red). The rates in units of $M_\odot$ per year (left panel) are seen to rise with increasing $z$, while the rates in units of per redshift (right panel) is largely independent of $z$. We also plot the total merger rate (haloes $+$ diffues) with dotted lines and overlay the fit presented in equation \ref{Mdotfit} in black.  Note that the fit is quite good and the total rate and fit are difficult to distinguish.} \label{fig:BCZ} 
\end{figure*}

Having examined the distributions of the growth rates at $z=0$ in the last section, we now study the redshift dependence of the {\it mean} growth rates. The left panel of Fig.~\ref{fig:BCZ} shows the mean $\B$ (solid), $\C$ (dashed), and total growth rate $\B+\C$ (dotted) as a function of redshift for three mass bins: $10^{12} M_\odot$ (blue), $10^{13} M_\odot$ (green), and $10^{14} M_\odot$ (red), where the threshold $\ximin=0.04$ is used to define $\B$ and $\C$. The right panel shows the same information except the rates are expressed in the dimensionless units of $dM/dz/M_0$. The growth rates per year (left) are seen to increase rapidly with increasing $z$, while the rate per unit $z$ (right) has a very weak dependence on $z$ out to $z\sim 6$. As a function of halo mass $M_0$, the right panel shows that the haloes of higher mass experience somewhat higher {\it fractional} mass growth rates than haloes of lower mass. Similar dependence on time and mass was seen for the halo merger rates in \citet{FM08}.

The left panel of Fig.~\ref{fig:BCZ} can be compared directly with the total mass growth rates shown in Fig.~4 of \cite{MFM09}, which is well approximated by equation~(7) proposed there: 
\begin{eqnarray}
	\left< \dot{M}_{\rm tot} \right> &=& 42 \, M_\odot {\rm yr}^{-1} \left( \frac{M}{10^{12} M_\odot} \right)^{1.127} \nonumber\\
	&& \times (1 + 1.17 z) \sqrt{\Omega_m (1+z)^3 + \Omega_\Lambda} \,. \label{Mdotfit} 
\end{eqnarray}
where $\Omega_m$ and $\Omega_\Lambda$ are the present-day density parameters in matter and the cosmological constant, and we have assumed $\Omega_m + \Omega_\Lambda=1$ (used in the Millennium simulation). We have overlaid this fit using black dotted curves in Fig.~\ref{fig:BCZ}. A comparison of the dashed and solid curves in Fig.~\ref{fig:BCZ} shows that the mean $\C$ is consistently $\sim 30$\% higher than the mean $\B$ at all mass and redshifts. The functional form of the mass and redshift dependence in equation~(\ref{Mdotfit}) is therefore also applicable for these separate rates; only the overall amplitude needs to be adjusted to obtain a fitting form for $\B$ and $\C$.

The similarity in shape and mass dependence of the $\B$ and $\C$ curves in Fig.~\ref{fig:BCZ} may lead one to conclude that the "non-halo" material contributing to $\C$ can be simply attributed to mergers with sub-resolution haloes. This interpretation is too simplistic, however, as we will discuss in Sections~4 and 5 below where their environmental dependence is analysed.

\section{The Environmental Dependence of Halo Growth Rates and Histories}

\subsection{Halo Mass Growth Rate due to Mergers vs Diffuse Accretion} \label{EnvDepofHGR}

\begin{figure*}
	\centering 
	\includegraphics{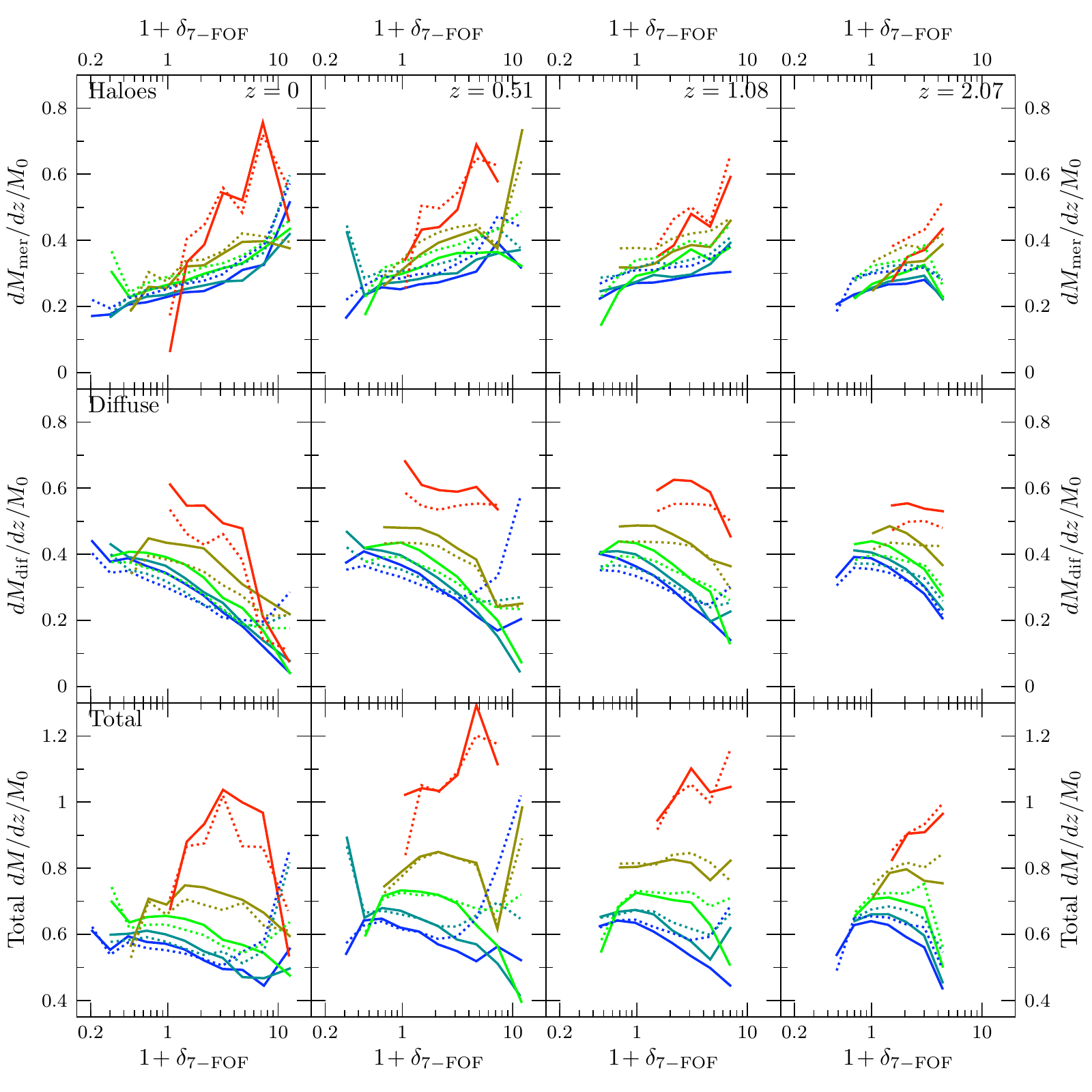} \caption{Environmental dependence of the halo mass growth rates due to contributions from merging progenitor haloes (top), diffuse accretion of non-halo mass (middle), and sum of the two (bottom). Four redshifts $z=\ZA,\ZB,\ZC,\ZD$ are shown for comparison (left to right). Within each panel, the five curves show five mass bins for the descendant haloes listed in Table \ref{table:MassBins} (blue: low mass, red: high mass) with solid lines computed using the subhalo mass definition and dashed lines computed using the FOF mass definition. On average, haloes in denser regions experience a higher mass growth rate from mergers and a lower rate from diffuse accretion than similar mass haloes in the voids. These two opposite trends with environment roughly cancel out to yield a weak, but net negative, dependence on $1+\dsfof$ for the total growth rate.} \label{fig:Instantaneous} 
\end{figure*}

In Part I of this series \citep{FM09}, we reported a strong positive correlation between the {\it number} of halo mergers and halo environment. We now turn to the environmental dependence of the rate of halo {\it mass} growth. Since haloes acquire mass through both mergers and diffuse accretion in the form of non-halo mass or unresolved haloes, we examine separately the two growth components defined in equation~(\ref{eqn:BC}).

The minimum mass ratio $\xi_{min}=0.04$ discussed in \S\ref{QuantifyingHaloGrowth} is used as the threshold between $\B$ and $\C$. We compute $\B$ and $\C$ for each descendant halo and then bin the results by averaging over various ranges of mass and $\delta$. For haloes with only one progenitor (above the mass threshold), there is no contribution from mergers by definition, and $\B=0$. These events are included when averaging over $\B$ so that a fair comparison can be made to the mass growth rate due to diffuse accretion, $\C$.

Fig.~\ref{fig:Instantaneous} compares the environmental dependence of the mean halo mass growth rate (per unit redshift) due to mergers (top row), diffuse accretion (middle), and the total rate (bottom) computed from the Millennium simulation. The rates at four redshifts are shown: $z=\ZA, \ZB, \ZC$ and $\ZD$ (from left to right). Within each panel, different colours correspond to different mass bins (listed in Table~\ref{table:MassBins}), where more massive haloes have higher growth rates. The solid curves are computed using $M_{SH}$, the sum of all the subhalo masses in a given FOF halo, while the dotted curves are computed using $M_{FOF}$ (see Sec.~2.3). Both mass definitions yield similar quantitative behaviour, though the subhalo mass definition yields slightly lower merger growth rates in accordance with the results in \citet{Genel}.

The mass growth due to mergers (top row) is seen to increase in denser regions, in agreement with the higher merger rates in denser regions reported in FM09. This is to be expected as the merger rate $B/n$ is related to the mass growth due to mergers $\B$ by 
\begin{equation}
	\B \approx \int_{0.04}^1 M_0\xi \frac{B}{n} \, d\xi \,.
\end{equation}
As with $B/n$, the environmental dependence of $\B$ persists out to higher redshift. 

By contrast, the middle row of Fig.~\ref{fig:Instantaneous} shows that the mass growth rate due to non-halo material has an opposite dependence on $\delta$: the mean value of $\CZ$ \emph{decreases} with increasing $\delta$ for all halo masses at all redshifts. This trend is clean when $M_{SH}$ is used as the mass definition (solid curves) but the curves show a sharp rise for low mass haloes when $M_{FOF}$ is used (blue dotted curve). We believe this is due to the inadequacy of the FOF halo finder in dense regions. As \cite{Genel} have shown, the FOF mass for low mass haloes sometimes rise as they approach high mass haloes; this may contribute to the rise in FOF mass observed in the densest regions. Despite this fact, we emphasise that the discrepancies between the $M_{FOF}$ rates and the $M_{SH}$ rates only arise in the densest regions (beyond $1+\dsfof \sim 5$). Outside of these regions, $\BZ$ still increases with $\delta$ and $\CZ$ decreases with $\delta$ regardless of the halo mass definition.

The mean total mass growth rate is shown in the bottom row of Fig.~\ref{fig:Instantaneous}. It shows that the strong but opposite dependence on environment found for the two growth rates act against each other, producing a relatively weak $\delta$-dependence for the overall growth rate that declines with increasing $\delta$ for the three low mass bins. This dependence is not as clearly seen for the very highest mass bin, though the statistics are poor for the cluster-scale bin (red curves). Again, the trends are much cleaner for $M_{SH}$ than for $M_{FOF}$ in the high-$\delta$ regions. The mass dependence of the total growth rates, on the other hand, is more prominent than the environmental dependence, with massive haloes growing more rapidly than low mass haloes. Equation~(7) of \citet{MFM09} (repeated above in eq.~(\ref{Mdotfit}) shows that this mass dependence is well approximated by a power-law: $\dot{M_{\rm tot}}\propto M^{1.127}$.

The curves in Fig.~\ref{fig:Instantaneous} are sensitive to the choice of cut used to remove the aberrant haloes (see Sec.~3.2) at the 20-50\% level. The qualitative features of the plot, including the markedly opposite correlation of the two rates with $\delta$, and the negative, but relatively weak, $\delta$ dependence of the total rate, are insensitive to choices of $f_{+}$ and $f_{-}$ in the cut and persist even if the cut is made asymmetric ($f_{+}\neq f_{-}$).

It should be pointed out that although the two growth rates depend on the choice of $\xi_{min}$ used to define resolvable haloes and diffuse material (0.04 in this case), the total growth rate is \emph{independent} of this parameter. Thus, redistributing progenitors with $\xi_{min}<0.04$ from $\CZ$ to $\BZ$ would not impact the bottom row of Fig.~\ref{fig:Instantaneous}. We will return to this point in more detail in \S\ref{InterpretingC}.

\subsection{Fraction of Final Halo Mass Acquired from Mergers vs Diffuse Accretion} \label{EnvDepofHGH}
\begin{figure}
	\centering 
	\includegraphics{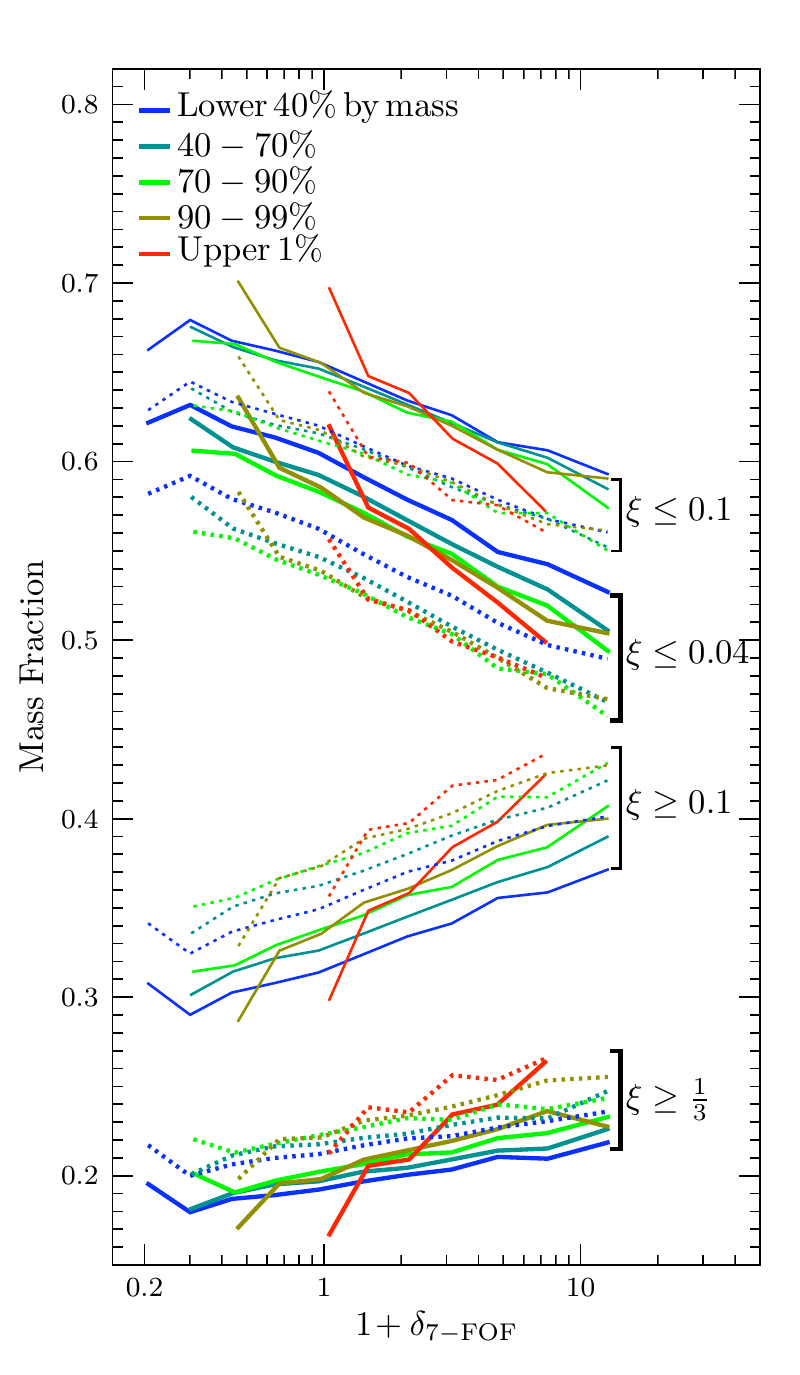} \caption{Environmental dependence of the fraction of the final halo mass $M_0$ gained by mergers vs diffuse accretion. The bottom two sets of curves show the rising mass fraction with $1+\dsfof$ for mass gained via mergers with progenitors of mass ratios $\xi>1/3$ and $\xi>0.1$. The top two sets of curves show the declining mass fraction with $1+\dsfof$ for the contributions from non-halo mass with $\xi<0.04$ and $\xi<0.1$. Within each set of curves, different colors represent different mass bins, and solid and dotted curves are for the two halo mass definitions $M_{SH}$ and $M_{FOF}$, respectively. This figure illustrates the increasing importance of mergers to halo growth in overdense regions and importance of diffuse accretion in the voids. } \label{fig:IC} 
\end{figure}

In addition to the \emph{instantaneous} halo mass growth rates shown in Fig.~\ref{fig:Instantaneous}, a related quantity of interest is the {\it integrated} contribution to a halo's final mass due to mergers vs diffuse accretion. Fig.~\ref{fig:IC} compares these contributions as a function of environment. The lowest set of curves plots the fraction of a halo's final mass gained from major mergers (defined to have progenitor mass ratio $\xi>1/3$ at the time of merger) through the halo's entire history (out to $z=6$); the set of curves above it shows the same quantity for more minor mergers with mass ratio $\xi > 1/10$. The five curves in each set show the five halo mass bins listed in Table~\ref{table:MassBins}. Similar to the top panels of Fig.~\ref{fig:Instantaneous}, these two sets of curves exhibit a {\it positive} environmental dependence, where haloes in denser regions experience more mergers and have a higher fraction of their mass coming in from mergers. From voids to overdense regions, the average mass fraction of a halo due to major mergers ($\xi>1/3$) increases from $\sim 20$\% to 25\%, and the mass fraction due to mergers with $\xi > 1/10$ increases from $\sim 30$\% to 40\%.

The upper two sets of curves show the analogous quantity for the fraction of a halo's final mass gained from material with $\xi<0.04$ and $\xi<0.1$. In contrast to the lower two sets of curves, the correlation of the mass contribution by non-halo matter with $\delta$ has the opposite sign, again consistent with the middle panels of Fig.~\ref{fig:Instantaneous}.

\begin{figure*}
	\centering 
	\includegraphics{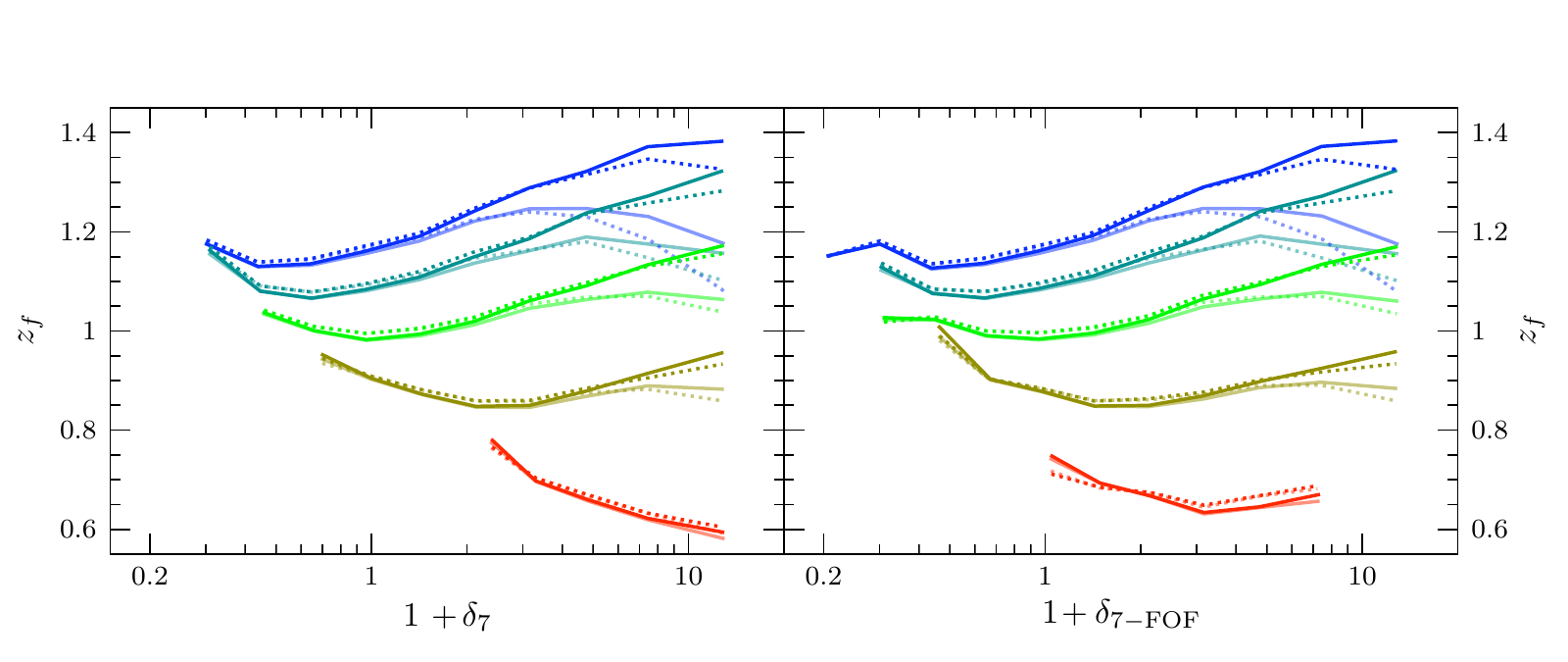} \caption{Environmental dependence of the mean formation redshifts for haloes in five mass percentile bins (same bins as Fig.~\ref{fig:IC}) in increasing mass from top down. Two measures of environment are shown for comparison: $1+\delta_7$ that includes the halo mass (left), and $1+\delta_{7-FOF}$ that excludes the central halo. Solid lines are computed using $M_{SH}$ and dotted lines using $M_{FOF}$. The darker colors are computed after the cut described in Section~\ref{BCDistribution} is applied; the light colors are computed without the cut. This figure shows that more massive haloes (bottom curves) on average form more recently. Within a mass bin, galaxy-size haloes in denser regions form earlier than those in the voids, while cluster-size haloes show little environmental dependence in their formation time.} \label{fig:ZF} 
\end{figure*}

A consistent picture is therefore emerging in that haloes in denser regions experience a {\it higher} rate of growth via mergers and a {\it lower} rate of growth via diffuse accretion than haloes of comparable mass in less dense regions. As a result, in dense regions, a higher fraction of a halo's final mass is gained through mergers than haloes in the voids. By contrast, diffuse accretion contributes less to halo mass growth in denser regions than in the voids. Overall, however, diffuse accretion (defined as $\xi < 0.04$) is an important component of a halo's final mass regardless of the environment: its contribution to the mass fraction is never lower than $\sim 40$\% (second set of curves from the top in Fig.~\ref{fig:IC}).

\subsection{Halo Formation Redshifts} \label{FormationRedshift}

We follow the convention in the literature and define the formation redshift $z_f$ for a halo to be the redshift at which the halo's mass first reaches $M_0/2$ (when traced backwards in time), where $M_0$ is measured at $z=0$. This is computed by tracing the mass of each of the $\sim 500,000$ haloes from low to high redshift in the Millennium outputs. A complementary $z_f$ can be defined by going from high to low redshift; we have done this as well and found little difference in our results.

Fig.~\ref{fig:ZF} shows the mean formation redshift for haloes in five mass bins (in increasing mass from top down) as a function of their environment at $z=0$. The two panels compare two definitions of the local densities discussed in \citet{FM09}: $1+\dsfof$ used throughout this paper (right) and the simpler $1+\delta_7$ (left). The two measures of local densities give very similar results except in the most massive halo bin (red curves), where $z_f$ shows little correlation with $\dsfof$ but {\it decreases} with increasing $\ds$. The negative correlation of $z_f$ with $\ds$ should not be interpreted as a true dependence on a halo's surrounding environment; instead, it is due to the tight correlation between $\ds$ and the masses of massive haloes shown in Fig.~2 of FM09 and the fact that, on average, more massive haloes form earlier than less massive haloes. This degeneracy between mass and density of massive haloes is removed when the variable $\dsfof$ is used (see Fig.~2 of FM09). The resulting $z_f$ vs $\dsfof$ shows almost no correlation in the right panel of Fig.~\ref{fig:ZF}.

The solid and dotted curves in Fig.~\ref{fig:ZF} compare the mean values of $z_f$ computed using the $M_{SH}$ vs $M_{FOF}$ definition of halo mass. As seen in Fig.~\ref{fig:Instantaneous}, the two mass definitions yield similar results in underdense regions but differ in the densest regions. Fig.~\ref{fig:ZF} also compares $z_f$ computed with (darker shades) and without (lighter shades) the cut described in Section~\ref{BCDistribution}. The two $z_f$'s are identical in underdense regions, but the sample without the cut has a lower $z_f$ at high $\delta$. This drop is unphysical and is due to the orhpaned halo fragments that suddenly appear at low $z$ and thus bias $z_f$ to artificially low values.

A number of conclusions can be drawn from Fig.~\ref{fig:ZF}. First, within a given environment, higher mass haloes on average form later (i.e. lower $z_f$). For instance, the mean formation redshift drops from $z_f\sim 1.1$ to 1.4 for $M \sim 10^{12} M_\odot$ haloes, to about 0.7 for $M > 10^{14} M_\odot$ haloes. This is a well known property of hierarchical cosmological models such as the $\Lambda$CDM. This mass dependence is also consistent with that shown in the bottom panels of Fig.~\ref{fig:Instantaneous}, where more massive haloes have a larger mean mass growth rate and hence a later mean formation time.

The second feature to note is that galaxy-size haloes (top few curves) in denser regions form earlier than the same mass haloes in the voids, whereas $z_f$ for cluster-size haloes (bottom curve) hardly correlates with environment (when $\dsfof$ is used as a measure). The earlier formation of galaxy haloes in denser regions is consistent with the halo assembly bias discussed in a number of recent papers \citep{Gao05, ShethTormen04, JingSutoMo07, GaoWhite07, Harker06, Wechsler06, Wetzel, WangMoJing07, Hahn08, MFM09}. Interestingly, \cite{Wechsler06} and \cite{JingSutoMo07} reported decreasing $z_f$ with increasing halo bias for $10^{13}-10^{14} M_\odot$ haloes. This trend is similar to the trend for the cluster-mass haloes (red curve) in the left panel of our Fig.~\ref{fig:ZF}. Since the connection between $\delta$ and halo bias is complex, it is not immediately clear that their results are inconsistent with ours. Perhaps the negative correlation observed by these authors is also due to the fact that the bias of high mass haloes is strongly correlated with mass and does not provide an independent measure of environment.

The third feature to note in Fig.~\ref{fig:ZF} is the slight upturn of $z_f$ in the most underdense void region. This trend makes intuitive sense since haloes in sufficiently underdense regions are deprived of fuel for growth, thereby growing more slowly and having larger formation redshifts. This trend does not seem to appear in any of the earlier literature except \cite{DesJacques07}, which noticed a similar trend in \cite{Harker06}. We suspect that the use of bias as a measure of halo environment in most of these earlier papers (as opposed to the local density used here and in \citealt{Harker06}) is not a sensitive probe of the $z_f$ statistics of haloes in the voids.

\subsection{Mass Reservoir outside Haloes} \label{ContributionsToDelta} 
\begin{figure}
	\centering 
	\includegraphics{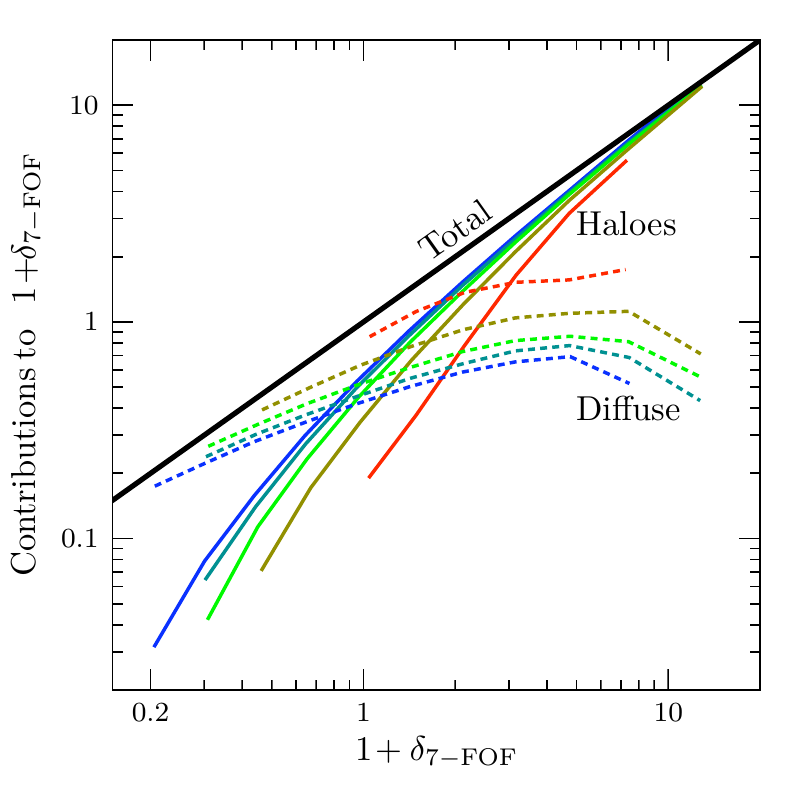} \caption{Composition of the mass reservoir in the region outside of the halo but within a 7 Mpc sphere centered at the halo. The vertical axis shows the contribution to $1+\dsfof$ by haloes (with mass above 4\% of the central halo; solid curves) and the remaining material (dashed lines). By construction, the solid and dashed curves add up to the thick diagonal line. Different colours denote the same mass bins in earlier figures. The mass reservoir for haloes in overdense regions is seen to be dominated by resolved haloes, whereas the reservoir for haloes in the voids is dominated by non-halo material. }
	\label{fig:Contributions} 
\end{figure}

In this subsection, we address the question: are the opposite environmental correlations of $\B$ and $\C$ with $\delta$ seen in Figs.~\ref{fig:Instantaneous} and \ref{fig:IC} also present in the surrounding mass reservoirs outside of the haloes? If yes, then the different environmental dependence of the growth rates is originated from the mass reservoirs that are feeding the haloes. If not, then different accretion timescales have to be operating for the haloes and non-halo material in overdense vs underdense regions. The answer to this question will therefore help explain the {\it origin} of the difference in the environmental correlations reported thus far.

To answer this question, we analyse the halo vs non-halo composition of the mass in the volume used to compute $\dsfof$ for each halo at $z=0$. Centered at each halo, we compute the mass within the 7 Mpc sphere (excluding the FOF mass itself) that resides in haloes with mass ratios exceeding $\xi>0.04$ (relative to the central FOF mass) vs in all other matter. The results are shown in Fig.~\ref{fig:Contributions}, which plots the mean values of these two components in the mass reservoir as a function of $1+\dsfof$ for five mass bins. The black diagonal line marks the sum of the two components, which by definition is simply $1 + \dsfof$.

Both the halo and non-halo components in the 7 Mpc reservoir outside the haloes are seen to increase with $1+\dsfof$, but the halo component has a much steeper slope. The two components contribute a comparable amount for haloes in the cosmic mean density (i.e. $\dsfof \sim 0$), but up to 80\% of the mass reservoir is made of haloes in very dense regions, while this fraction drops to $\sim 20$\% in voids, where the diffuse component dominates. Even though only a fraction of this mass reservoir is accreted onto the central haloes and contributes to actual halo growths, Fig.~\ref{fig:Contributions} suggests that the higher merger rates in denser environments are a direct consequence of the larger fraction of available haloes in the reservoir.

A somewhat subtle but important feature to note in Fig.~\ref{fig:Contributions} is that the \emph{absolute} amount of diffuse mass in the reservoir \emph{rises} with increasing $\dsfof$ even though the {\it fractional} contribution due to this component drops. This rising trend with $\dsfof$ of the diffuse component is in striking contrast to the middle panels of Fig.~\ref{fig:Instantaneous}, where the accretion rates of diffuse material, $\C$, onto haloes are seen to {\it decrease} in denser regions. The naive interpretation that $\C$ decreases with increasing $\dsfof$ is due to a dwindling reservoir of diffuse material outside of the haloes in denser regions is therefore invalid. On the contrary, at a fixed halo mass, there is more non-halo material available in the regions where $\C$ is lower.

Additional gravitational effects are likely to be at work in dense
environments to explain why the halo mass growth rate due to diffuse
material is stunted in denser regions while the supply of this mass is in
fact larger.  For instance, the non-halo component in the reservoir may be
dynamically hotter than the resolved haloes due to tidal heating and
stripping, thereby reducing $\C$ in higher density regions as shown in
Fig.~\ref{fig:Instantaneous}. Pieces of this picture have been suggested in
the literature. \cite{WangMoJing07} noted that a halo's ability to consume
the available fuel is as important as the amount of fuel available for
growth. They found that haloes in dense regions accreted matter more slowly
than expected, although they did not examine the separate $\B$ and $\C$
components as we do here.  It has also been suggested that low-mass haloes
form earlier because their late accretion is eliminated in a competition
with massive neighbors \citep{ZentnerEPS}. Fig.~\ref{fig:Instantaneous},
however, shows that this competition idea is not strictly true: regardless
of mass, all haloes in overdense regions have both increased growth rate
$\B$ and decreased growth rate $\C$. The growth rates of low mass haloes
are therefore not simply controlled by competition with nearby high-mass
haloes in overdense regions.  Alternatively, the diffuse material
surrounding very dense regions may preferentially accrete onto the
lower-density filaments and the haloes within them, which then infall onto
the clusters.  Further analysis taking into account of filaments as well haloes
would be required to test this idea.

\subsection{Time Evolution of a Halo's Environment} \label{Sojourners} 

Thus far we have used the mass densities computed at a halo's redshift to
quantify its environment. Here we examine whether the environments of
haloes evolve significantly over their lifetimes, that is, if the haloes in
overdense regions today tended to have progenitors that also resided in
overdense regions at earlier times.

To study such environmental evolution, we first group the haloes at $z=0$
into mass bins. Within each mass bin, the haloes are further divided into
five environmental bins, each corresponding to 20 percentile in the
distribution of $1+\dsfof$. Each halo is therefore assigned to a mass and a
$\dsfof$ bin. For each halo at $z=0$ we then identify its most massive
progenitor at $z=\ZB$ and $z=\ZC$. We assign each progenitor a mass bin at
its redshift and compute the $\dsfof$ bin to which it belongs, using the
density field at that given $z$.

We then compute the fraction of haloes at $z=0$, as a function of mass bin,
whose most massive progenitors reside in the \emph{same} percentile
$\delta$ bin at $z=\ZB$ and $\ZC$. Such haloes have not deviated from their
local environment significantly over time. Since $\delta$ evolves with
time, it is more meaningful to compare relative percentiles instead of
absolute values of $\delta$ at different redshifts. We also compute the
fraction of haloes at $z=0$ that are at most one percentile bin away at
$z=\ZB$ and $\ZC$ . These haloes have moved slightly outside of their $z=0$
environmental percentile.

We find that $\sim 70$\% of $M \la 10^{13} M_\odot$ haloes do not change
environmental bins between $z=\ZB$ and today, and 60\% do not change
environments between $z=\ZC$ and today.  In addition, more than 95\% of
these haloes are within one $\delta$-percentile bin of their final
environment.  The range of $\delta$ for cluster-mass haloes is too narrow
(see, e.g., Fig.~6) for this analysis to be meaningful.  Thus, the majority
of haloes reside in the same environmental context across their lifetimes.

\section{Discussion} \label{Discussion}

\subsection{Interpreting ``Diffuse'' Accretion }
\label{InterpretingC}

In Sections~3 and 4, we presented results for the mass accretion rates of
dark matter haloes through mergers ($\B$) and diffuse non-halo material
($\C$), and their respective environmental dependence. As emphasised there,
merging progenitors with a mass ratio of $\xi>\ximin=0.04$ were counted
towards $\B$, whereas the rest of the mass growth was counted towards $\C$.

\begin{figure}
	\centering 
	\includegraphics{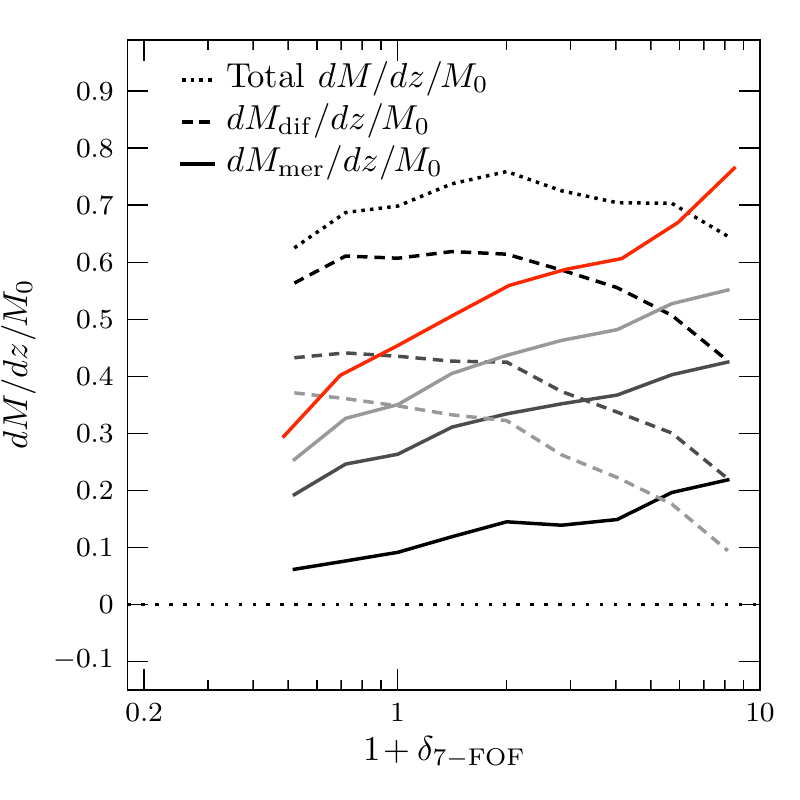} \caption{Dependence of the two growth
          rates $\BZ$ (solid) and $\CZ$ (dashed) on the mass threshold
          $\ximin$ used to define the halo and non-halo components. The sum
          of the two rates (dotted), by construction, is independent of
          $\ximin$. Haloes in the $90-99\%$ mass bin are used. As $\ximin$
          is lowered from 0.4 (black), 0.04 (dark grey), to 0.004 (light
          grey), the amplitude of $\BZ$ increases and that of $\CZ$
          decreases, but their respective correlations with $\dsfof$ remain
          the same, suggesting that the environmental dependences reported
          in this paper are robust and do not depend on the exact values of
          $\ximin$ used to define haloes vs non-haloes. The top solid curve
          (in red) shows the hypothetical value of $\BZ$ extrapolated down
          to $\ximin=0$ (see text for discussion). One explanation for the
          $\delta$-dependent gap between this curve and the total rate
          (dotted) is that there exists a truly diffuse component of dark
          matter that does not belong to bound haloes, in particular in the
          lower-density regions.} \label{fig:Resolution}
\end{figure}

To test the robustness of the results in Sections~3 and 4 with respect to
the value of $\ximin$ used to define haloes vs non-haloes, we take haloes
in one of the more massive bins (the 90-99\% mass bin, corresponding to
$1.4\times 10^{13}$ to $1.1 \times 10^{14} M_\odot$) that provides good
statistics as well as high mass dynamic range, and recompute the two rates
using different values of $\ximin$. The results are shown in
Fig.~\ref{fig:Resolution}, which plots the environmental dependence of the
two mass growth rates for three values of $\ximin$: 0.4 (black), 0.04 (dark
grey; the value used in earlier figures), and 0.004 (light grey). As
$\ximin$ is lowered, more of the mass is counted towards the merger
component, so $\BZ$ (solid curves) increases and $\CZ$ (dashed curves)
decreases, while the sum of the two (dotted curve) is, by construction,
independent of $\ximin$. We note that even though the overall amplitudes of
the two rates change with $\ximin$, $\BZ$ remains positively correlated and
$\CZ$ remains negatively correlated with $\dsfof$ regardless of
$\ximin$. This test suggests that the environmental correlations shown in
Figs.~\ref{fig:Instantaneous} and \ref{fig:IC} are robust to the value of
$\ximin$ used to define haloes vs non-haloes.

We explore further the nature of the diffuse non-halo component by
examining the hypothesis that this component is comprised entirely of
sub-resolution haloes such that in the limit of $\ximin \rightarrow 0$ (or
more precisely, the true minimal halo mass, which is likely to be about an
Earth mass \citep{Diemand05}, but the difference is negligible), we would
expect $\C\rightarrow 0 $ and $\B\rightarrow M_0-M_1$. To assess whether
this hypothesis is plausible, we recompute $\B$ by extrapolating down to
$\ximin=0$, assuming that $\B$ maintains the same power-law dependence on
$\xi$ determined for a given mass bin in the simulation.\footnote{We note
  that our best-fit slope for the $\xi$-dependence of $\B$ in \citet{FM08}
  was in fact slightly steeper than $-2$, which would give a divergent
  growth rate. Our goal there was to obtain a simple universal form. For
  better accuracy, here we fit $\B$ vs $\xi$ for each mass bin
  independently, where the slope had a very mild mass dependence, ranging
  from $-1.7$ to $-1.9$.}

The predicted mass growth rate under this assumption is plotted as a
function of environment in Fig.~\ref{fig:Resolution} (top solid red curve).
In comparison with the total rate $\B+\C$ (dotted curve) determined from
the Millennium simulation, we observe a clear environment-dependent gap
between these two curves that widens in the lower-density regions. This
gap suggests two possible scenarios: (1) the power-law behaviour of $\B$
steepens towards low $\xi$ (an effect that is not observed down to $\xi\sim
10^{-4}$ resolvable by current simulations), or (2) there exists a truly
diffuse component of dark matter that does not belong to bound
haloes. Moreover, both scenarios must be dependent on environment. For (1),
the $\delta$-dependent gap in Fig.~\ref{fig:Resolution} can only be closed
if the power-law slope of $\B$ vs $\xi$ for sub-resolution progenitors is a
strong function of halo environment (being steeper in underdense
regions). We cannot rule out this possibility without testing it in
higher-resolution simulations, but all the halo properties that we have
examined thus far in FM08 and FM09 do not exhibit such complex,
non-universal behavior. On the contrary, down to the current simulation
resolution, the halo-halo merger rates and mass growth rates as a function
of progenitor mass ratio $\xi$ were all well fit by a single power law,
where the slope was entirely independent of the halo environment (see
Figs.~4, 5, and 7 of FM09). We therefore favour scenario (2) that the
diffuse non-halo mass has a truly diffuse component, and this component is
more prominent in lower-density regions. New simulations with higher mass
resolution such as the Millennium-II \citep{BK09} will be used to explore
further these two possibilities.

\subsection{Implications for the Extended Press-Schechter and Excursion Set Models} \label{EPS}

The Press-Schechter model \citep{PS74} assumes that primordial
Gaussian-distributed dark matter perturbations collapse into haloes of mass
$M$ when their linearly extrapolated overdensities, smoothed on a scale
corresponding to a given mass, exceed a critical threshold computed using a
simple spherical collapse model.

This framework was extended to address the problem of halo growth in the
excursion set formalism (sometimes called the Extended Press-Schechter or
EPS model) \citep{BondEPS, LC93}. Using a $k$-space tophat smoothing
window, these authors showed that the density perturbation at a given point
in space undergoes a random walk as the perturbation is smoothed on smaller
and smaller scales. These perturbation ``trajectories'' obey the diffusion
equation and are Markovian, i.e., the change in overdensity at a given
scale is independent of the overdensities at other scales.
The EPS model in its standard form therefore predicts no environmental
dependence. For example, the conditional mass function
$\phi(M_1,z_1|M_0,z_0)$ -- the distribution of progenitor mass $M_1$ at
redshift $z_1$ for a descendant halo of mass $M_0$ at $z_0$ -- is predicted
to be the same for all environments, in contrast to Fig.~7 of \citet{FM09}
that showed descendant haloes in dense regions to have more progenitors
than those in voids. 

Recent attempts have been made to introduce environmental correlations into
the EPS model. As emphasised in \cite{ZentnerEPS}, the Markovian nature of
the random walks in the excursion set model is \emph{not} a prediction but
is an assumption originating from the use of the $k$-space tophat window
function. This window function is traditionally chosen to reduce the time
required for computing density perturbation trajectories since for other
window functions, the perturbations must be computed at all scales
simultaneously. When a Gaussian window function was used, \cite{ZentnerEPS}
indeed found an environmental dependence of the halo formation redshift,
but the dependence was {\it opposite} to that seen in numerical simulations
(including Fig.~\ref{fig:ZF} of this paper), where older (i.e. earlier
forming) haloes tended to be more clustered.

\cite{DesJacques07} instead chose to introduce environmental dependence
into the threshold density for collapse in the ellipsoidal excursion set
formalism. Overdense regions will tend to exert more tidal shear on
collapsing haloes, an effect that causes the haloes to virialize
earlier. Like \cite{ZentnerEPS}, however, \cite{DesJacques07} found the low
mass haloes in denser regions to form later than similar mass haloes in
emptier regions, again opposite to the effect seen in
simulations. \cite{DesJacques07} did note that his results may be valid in
the lowest-density region, where Fig.~\ref{fig:ZF} here and \cite{Harker06}
both reported a subtle increase in formation redshift with decreasing
$\delta$.

\cite{Sandvik07} carried out a multi-dimensional generalisation of EPS that
allowed them to incorporate environmental effects by tracking the shapes of
collapsing haloes and incorporating information about gravitational
shear. They found $z_f$ in this approach to depend very weakly on the
environment; moreover, the $\delta$ dependence they did observe was
stronger for more massive haloes, in contrary to the results of N-body
simulations such as our Fig.~\ref{fig:ZF}.  They proposed, instead, that the
environmental dependence of a halo's formation history may be related to
the halo's progenitor pancakes and filaments. They found that haloes whose
progenitors were the most massive pancakes or filaments were more clustered
than halo's whose progenitors were the least massive pancakes or filaments,
and that this clustering dependence was stronger for \emph{low mass}
haloes.

In summary, there is not yet a modification of the EPS model that can
successfully predict the correct correlation of halo formation with
environment seen in numerical simulations.  We emphasize that in addition to
the ``assembly bias'' for the formation redshift discussed in Sec~4.3 and
shown in Fig.~\ref{fig:ZF}, an improved EPS model would also need to
predict the strong but opposite environmental dependence of the growth
rates shown in Fig.~\ref{fig:Instantaneous}.  A first step towards this goal
would be to introduce into the model, and track, the contribution made by
the non-halo material $\C$.  The model would also need to be sophisticated enough to account for the
opposite behaviours of $\B$ and $\C$ as a function of the local density.

\section{Conclusions} \label{Conclusions}

We have used the Millennium simulation \citep{Springel05} to investigate
the dependence of halo mass growth rates and histories on halo
environment. This paper complements our previous paper \citep{FM09} in
which we reported and quantified how haloes in overdense regions
experienced higher merger rates than haloes in underdense regions. Here, we
have studied the mass growth rates due to both mergers with other resolved
haloes, $\B$, and accretion from non-halo ``diffuse'' material,
$\C$. Results for the distributions of $\B$ and $\C$ at $z=0$ and the
redshift evolution of the mean $\B$ and $\C$ are summarized in
Figs.~\ref{fig:BCDistribution}-\ref{fig:BCZ}.

As a function of halo environment, we have found (see
Fig.~\ref{fig:Instantaneous}) that the growth rate $\B$ due to mergers, in
agreement with the merger rates in \citet{FM09}, is positively correlated
with the local density, whereas $\C$ is negatively
correlated. Consequently, in denser regions, mergers play a {\it
  relatively} more important role than diffuse accretion for the mass
growth of haloes, and a higher fraction of a halo's final mass is acquired
through mergers in these regions than in the voids (Fig.~\ref{fig:IC}).

We have shown that the origin of the environmental dependence of growth
rate due to mergers, $\B$, is directly linked to the mass reservoir
immediately outside the virial radii of the haloes (see
Fig.~\ref{fig:Contributions}).  The mass composition in these surrounding
regions exhibits a strong positive correlation with environment such that
more than 80\% of the mass is in the form of resolved haloes for haloes
residing in dense regions (for $1+\dsfof \ga 5$), while only $\sim 20$\% of
the mass is in resolved haloes for haloes residing in voids (for $1+\dsfof
\la 0.4$).  Even though only a fraction of these haloes enters the virial
radii and contributes to the actual growth of the central halo, it is the
environmental property of this mass reservoir that leads to the
environmental dependence of $\B$.  For the diffuse growth component $\C$,
however, its negative correlation with local density observed in
Fig.~\ref{fig:Instantaneous} is {\it not} explained by the environmental
dependence of the available diffuse mass in the reservoir outside of the
haloes.  In fact, more diffuse material is available in the reservoir in
denser regions (dashed curves in Fig.~\ref{fig:Contributions}).  Why then
is the diffuse accretion rate $\C$ {\it lower} in denser regions?  We
speculated that some of the diffuse mass may be dynamically hotter due to
tidal stripping and therefore is harder to accrete.  A careful analysis
using the actual particle data from the simulation would be needed to
investigate this issue further.

The halo growth rates $\B$ and $\C$ together account for the overall halo
mass accretion history (MAH) that we have studied in detail in a separate
paper \citep{MFM09}. When combined, the opposite correlations of the two
rates with the local density cancel each other to some extent, resulting in
a weak environmental dependence for the formation redshift $z_f$
(Fig.~\ref{fig:ZF}). At a fixed mass, galaxy-sized haloes in overdense
regions on average form earlier than those in underdense regions,
consistent with the assembly bias result reported in several recent papers
\citep{Gao05, ShethTormen04, JingSutoMo07, GaoWhite07, Harker06,
  Wechsler06, Wetzel, WangMoJing07, Hahn08}. The $z_f$ for cluster-sized
haloes, on the other hand, show no dependence on environment. We note that
had we ignored $\C$ and only taken into account the mass growth due to
mergers, then the positively-correlated $\delta$ dependence of $\B$ would
imply a later formation redshift (i.e., a smaller $z_f$) for haloes in
denser regions, which is opposite to that shown in Fig.~\ref{fig:ZF}. The
negatively correlated $\C$ with the local density therefore plays a crucial
role in counterbalancing the positively correlated $\B$ so that the
environmental dependence of the total rate is consistent with that of
$z_f$.

We have emphasized throughout the paper that the values of $\B$ vs. $\C$
depend on the mass threshold $\xi_{\rm min}$ used to defined these two
components. Despite this fact, our tests have shown (see
Fig.~\ref{fig:Resolution}) that the enviromental trends of $\B$ and $\C$ in
Figs.~\ref{fig:Instantaneous} -- \ref{fig:Contributions} are robust and are
independent of the definition.

A number of interesting questions remain to be explored. For instance, what
is the nature of the non-halo component $\C$? Besides tidal stripping, are
there additional physical processes controlling the different environmental
dependence of $\B$ vs $\C$? How do our results for $\B$ and $\C$ extend to
higher-resolution simulations, e.g., will the power-law slope of $\B$ vs
progenitor mass ratio $\xi$ remain the same or steepen at smaller $\xi$,
and will this change be $\delta$-dependent? How does one introduce
non-Markovian features into the random-walk picture in the much-used EPS
model in order to reproduce the halo growth histories as a function of halo
mass and halo environment reported in this paper? We expect to shed light
on some of these questions by using the Millennium-II simulation
\citep{BK09} with $\sim 1000$ times better mass resolution in an upcoming
paper.

\section*{Acknowledgements}

We thank the anonymous referee for insightful comments. This work is
supported in part by NSF grant AST 0407351.  The Millennium Simulation
databases used in this paper and the web application providing online
access to them were constructed as part of the activities of the German
Astrophysical Virtual Observatory.

\bibliographystyle{mn2e} 
\bibliography{FM09B}

\label{lastpage}
\end{document}